\documentclass[final,5p,times,twocolumn,authoryear]{elsarticle}

\usepackage{amssymb,amsmath}
\usepackage{csquotes}
\usepackage[hidelinks]{hyperref}
\hypersetup{
    colorlinks,
    linkcolor=blue,
    citecolor=blue,
    urlcolor=blue
}
\usepackage{balance}

\usepackage{algorithm}
\usepackage{algpseudocode}

\usepackage{booktabs}

\makeatletter
\renewcommand{\ALG@name}{Appendix}
\makeatother

\begin{document}

\begin{frontmatter}



\title{Towards a standardized methodology and dataset for evaluating LLM-based digital forensic timeline analysis}

\author[author1]{Hudan Studiawan\corref{cor1}}
\ead{hudan@its.ac.id}
\cortext[cor1]{Corresponding author}
\affiliation[author1]{organization={Department of Informatics, Institut Teknologi Sepuluh Nopember},
            country={Indonesia}}

\author[author2]{Frank Breitinger}
\ead{frank.breitinger@uni-a.de}
\affiliation[author2]{organization={Institute of Computer Science, University of Augsburg},
             city={Augsburg},
            country={Germany}}

\author[author3]{Mark Scanlon}
\ead{mark.scanlon@ucd.ie}
\affiliation[author3]{organization={Forensics and Security Research Group, School of Computer Science, University College Dublin},
          country={Ireland}}

\begin{abstract}

Large language models (LLMs) have seen widespread adoption in many domains including digital forensics. While prior research has largely centered on case studies and examples demonstrating how LLMs can assist forensic investigations, deeper explorations remain limited, i.e., a standardized approach for precise performance evaluations is lacking. Inspired by the NIST Computer Forensic Tool Testing Program, this paper proposes a standardized methodology to quantitatively evaluate the application of LLMs for digital forensic tasks, specifically in timeline analysis. The paper describes the components of the methodology, including the dataset, timeline generation, and ground truth development. Additionally, the paper recommends using BLEU and ROUGE metrics for the quantitative evaluation of LLMs through case studies or tasks involving timeline analysis. Experimental results using ChatGPT demonstrate that the proposed methodology can effectively evaluate LLM-based forensic timeline analysis. Finally, we discuss the limitations of applying LLMs to forensic timeline analysis.
\end{abstract}

\begin{keyword}
LLM evaluation \sep Forensic timeline \sep Large language models \sep ChatGPT \sep log2timeline/Plaso
\end{keyword}

\end{frontmatter}


\section{Introduction}
\label{sec:introduction}

Forensic investigations often require the reconstruction of a timeline of events and activities related to a digital device or users \citep{Hargreaves2012}. Such timelines can provide valuable insights into various criminal activities, including malware, brute-force attacks, or attacker post-exploitation activities. The timeline analysis process is complex and time-consuming, particularly when dealing with large amounts of digital data from multiple sources. Traditional methods for timeline analysis are based on manual analysis, which can be subjective and prone to errors, and can lead to missing critical information \citep{Studiawan2020}.

The development of large language models (LLMs), such as OpenAI's GPT-3 \citep{brown2020language} has opened up many possibilities, including in digital forensic research. The model has been implemented in the ChatGPT application and instantly gained many users \citep{Buchholz2023}. LLMs have shown remarkable performance in various natural language processing tasks, including language generation and question-answering. Leveraging these capabilities, natural language processing techniques can be applied to digital data sources to analyze temporal information and investigate timelines of events. Other studies also suggest that artificial intelligence should provide more assistance in forensic investigation \citep{hall2022explainable,Studiawan2019}.

An editorial article by \cite{scanlon2023digital} discusses the increasing demand for expert digital forensic analysts and the potential use of LLMs such as ChatGPT in this domain. They emphasize the importance of maintaining the ``AI-assisted investigation" and ``human-in-the-loop" mantras when using LLMs in digital forensics. The article suggests that LLMs could lead to a new career specialization of digital forensic prompt engineers. \citet{WICKRAMASEKARA2025301859} provides a comprehensive overview of where LLMs may assist digital forensics. 

In addition, various studies explored the application of LLMs in digital forensics. For instance, \citep{scanlon2023chatgpt} assessed ChatGPT's impact on tasks such as understanding artifacts, evidence searching, and anomaly detection. Although ChatGPT shows promise in several low-risk forensic applications, concerns arise about evidence security and the model's occasional inaccuracies. Experts must exercise caution and have a deep understanding of the subject to effectively use ChatGPT in forensic scenarios.
Furthermore, ChatGPT has been explored for digital evidence investigations \citep{henseler2023chatgpt}, virtual forensic assistants \citep{dinis2023chatgpt}, and report writing \citep{michelet2024chatgpt}. Based on our literature review, existing work in this area has not discussed standardized evaluation for LLM-based digital investigation. 

\textit{Contribution.} The contributions of this paper are as follows:
\begin{enumerate}
    \item This paper proposes a standardized methodology to quantitatively evaluate the performance of LLMs in forensic timeline analysis tasks, such as event summarization. 
    \item This study presents a case study of forensic timeline analysis using LLM, e.g., ChatGPT.
    \item We created forensic timeline datasets and ground truth from Windows 11 using Plaso and these are publicly available\footnote{\url{https://zenodo.org/blinded\_for\_review}} for research and education purposes.
\end{enumerate}

The remainder of the paper is organized as follows: Sec.~\ref{sec:related-work} provides related research. Sec.~\ref{sec:proposed-method} describes the proposed approach for standard methodology and quantitative evaluation for LLM-based timeline analysis. Sec.~\ref{sec:experiments} presents the case study that demonstrates the application of the proposed method and a discussion of the results. Finally, Sec.~\ref{sec:conclusion} concludes this study.

\section{Related work}\label{sec:related-work}

\subsection{Forensic tool testing and validation}

To effectively validate digital forensic tools and methods, a proper validation test plan should include laboratory use in the real world, controlled internal tests based on scientific principles, and peer review. \citet{brunty2023validation} provides an overview of the foundational scientific aspects of forensic validations and describes the recommended steps to conduct a forensically sound validation method.

The Computer Forensics Tool Testing (CFTT) Program at NIST aims to establish a methodology for testing computer forensic tools, including developing specifications, test procedures, and criteria \citep{nist2019computer}. The program helps to ensure the reliability of forensic software tools, helping tool makers, users, and interested parties. CFTT methodology involves breaking down forensic tasks into discrete functions and creating test methodologies for each. 

\citet{hughes2020towards} discuss the need for rigorous validation practices in digital forensics to establish accuracy and reliability. They highlight challenges in developing statistical confidence for forensic tools, such as the lack of reference data, validation methods, and precise definitions of measurement. The authors propose a method for generated data procedures, virtual machine-based validation, and empirical models to guide the analysis.

Another study discusses the challenges of scientifically validating digital forensic evidence \citep{arshad2018digital}. The authors emphasize the lack of standard datasets, formal testing procedures, and established error rates. \citet{horsman2019tool} examines the challenges of ensuring reliability in digital forensic tools. The paper discusses the lack of standardized validation methods and the issues of transparency from software vendors. A survey of practitioners reveals widespread concerns about tool reliability and a need for improved testing standards and error rate reporting.

The related study on tool testing and validation shows a research gap where we need a method to evaluate and validate LLMs as tools in digital forensics. This paper aims to fulfill this need specifically for the task of forensic timeline analysis.

\subsection{Forensic timeline analysis}

Forensic timeline analysis involves reconstructing the sequence of events and activities related to a user or a system. Therefore, a variety of artifacts, such as browsing history, log files, or file metadata, are being parsed, and relevant information is extracted \citep{palmbach2020artifacts}. The analysis of the timeline analysis is then conducted using tools and data visualization techniques \citep{Inglot2014}. If tools do not yield expected results, a manual examination of data sources may be required. However, this approach can be time-consuming, labor-intensive, and prone to errors.

Timeline generation tools, such as log2timeline/Plaso, Autopsy, and Magnet AXIOM, can automate the timeline analysis process to some extent by extracting relevant temporal information from digital data sources. However, these tools are limited by the quality of the extracted data and may not be able to capture all relevant events and activities from acquired artifacts \citep{studiawan2022forensic}.

The approach by \cite{Hargreaves2012} can automatically reconstruct or summarize high-level events from low-level events. Previous techniques focus on extracting times from a disk image into a timeline, which can produce several million ``low-level'' events (e.g., file modification or Registry key update) for a single disk. In contrast, this approach can automatically reconstruct high-level events (e.g. connection of a USB stick) from this set of low-level events. The knowledge representation model presented in \cite{chabot2014complete} allows a semantically rich representation of events related to the incident. It includes the identification of correlated events that can highlight valuable information for the investigators.

The construction of a timeline array using time information from web browser log files is one way to perform forensic timeline analysis \citep{nalawade2016forensic}. Different data types of timelines that can be constructed from web browser artifacts such as web history, cache, cookie, download history, and search term timelines. Furthermore, \cite{Bhandari2020} propose an abstraction-based approach to reconstruct a timeline of events and artifacts. The method enhances the relevance of the timeline by reconstructing it into four levels of depth, from general to specific, to reduce complexity and extract information. 

The use of deep learning techniques, e.g., autoencoders, improves anomaly detection in a forensic timeline by establishing a baseline for normal activities \citep{Studiawan2021}. Another tool, namely DroneTimeline, constructs a timeline from a drone device and considers time extracted not only from file metadata, but also from various source artifacts of a drone or its control devices \citep{studiawan2022dronetimeline}. 

\subsection{LLMs for digital forensics}

In the case of LLM application for digital forensics, \citet{henseler2023chatgpt} discuss how ChatGPT can assist  investigators by writing structured queries, summarizing and evaluating large volumes of communication data, and analyzing search results. The authors highlight that ChatGPT can transform natural language queries into structured formats, summarize and visualize chat logs to reveal key relationships. The study notes limitations, such as hallucinations and the need for expert guidance. Another work explores the potential of using LLMs, e.g., ChatGPT and Llama, to assist in the generation of forensic reports in digital investigations \citep{michelet2024chatgpt}. The authors assess the ability of LLMs to automate parts of the report writing process, focusing on sections such as the introduction, items received, methodology, and results. They found that while ChatGPT performs well and generates relatively accurate drafts, Llama struggles with accuracy and completeness. The results show that LLM outputs still require proofreading and corrections. 

\citet{dinis2023chatgpt} also explore the potential and challenges of using ChatGPT in forensic sciences. The authors highlight the advantages of ChatGPT, such as assisting forensic professionals in drafting reports, analyzing forensic data, performing literature searches, and serving as a virtual forensic assistant. However, the paper also raises concerns about the ethical and legal challenges associated with using AI in this field, such as credibility issues, inaccuracies, plagiarism, and the risk of overreliance on AI in judicial decisions. 

Finally, \citet{scanlon2023chatgpt} describes the potential applications of ChatGPT and LLMs in digital forensics. The authors assess how ChatGPT can assist in various forensic tasks, such as identifying digital artifacts, generating code for forensic activities, and detecting anomalies in logs. LLMs present challenges including issues with hallucinations, inaccuracies, and limitations when dealing with sensitive data. The study shows that ChatGPT can be a useful tool for investigators when used with caution, but human expertise remains essential to ensure reliability in forensic investigations.

The application of LLMs in digital forensics has the potential to enhance investigators' capabilities to handle digital evidence and help solve cases with greater accuracy. However, it is important to remember that LLMs are not a replacement for human expertise, but rather a valuable tool that complements and assists forensic professionals. Therefore, we need a methodology and a dataset to evaluate LLMs as a forensic tool, particularly for timeline analysis, as discussed in this paper.

\begin{figure}[!t]
	\centering
	\includegraphics[width=0.45\textwidth]{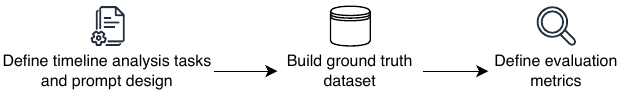}
	\caption{The proposed methodology for quantitative evaluation of LLM-based timeline analysis}
	\label{fig:proposed-method}
\end{figure}

\section{Proposed methodology}
\label{sec:proposed-method}

To assess the performance of an LLM for timeline analysis, several aspects are important as depicted in Fig.~\ref{fig:proposed-method}. 
We must define one or more tasks (Sec.~\ref{sec:task_definition}) that we expect the LLM to perform. This involves designing a prompt to interact with the system, such as summarizing events into high-level insights or identifying indicators of compromise.
In addition, a ground-truth dataset is needed that can be used to assess the outcome of an LLM (Sec.~\ref{sec:ground_truth}). Lastly, evaluation metrics are required that allow us to compare the ground truth with LLM output (Sec.~\ref{subsec:evaluation-metrics}). 
While starting with the tasks may seem natural, we recommend beginning with the evaluation metric, as it defines the required output, which in turn influences the task and prompt.

\subsection{Evaluation metrics}
\label{subsec:evaluation-metrics}
For the evaluation, we decided to use BLEU (Bilingual Evaluation Understudy) and ROUGE (Recall-Oriented Understudy for Gisting Evaluation). They were selected due to their widespread acceptance and established methodologies in machine translation and summarization. These metrics provide a way to quantify the quality of generated text and allow for comparisons across different models and tasks.

\subsubsection{BLEU -- Bilingual Evaluation Understudy}
BLEU assesses the quality of machine-generated outputs by comparing them to human-curated reference texts (ground truth) \citep{papineni2002bleu}. The score focuses on how accurately and completely the machine or LLM has replicated the human ground truth. It is calculated as follows:
\begin{equation}
    \text{BLEU} = \text{BP} \times \exp\left(\sum_{n=1}^{N} w_n \log p_n\right)
\end{equation}
where $p_n$ is the precision for each $n$-gram, $w_n$ is the weight for each $n$-gram, and $BP$ is the brevity penalty (BP). BP is designed to penalize generated text that is too short. The idea is that shorter text might artificially increase precision, but may not capture the full meaning of the original text. The brevity penalty is calculated as:
\begin{equation}
    \text{BP} = 
          \begin{cases} 
          1 & \text{if } c > r \\
          e^{(1-r/c)} & \text{if } c \leq r
          \end{cases}
\end{equation}
where $c$ is the length of the candidate (machine) translation and $r$ is the reference length.

\subsubsection{ROUGE -- Recall-Oriented Understudy for Gisting Evaluation}
ROUGE is a collection of metrics designed to evaluate automatic summarization and machine translation systems \citep{lin2004rouge}. It primarily focuses on the quality of the output generated by these systems. In our case, the essence of ROUGE is to provide a quantitative measurement of the quality of an automatically generated text from an LLM by comparing it with reference data or ground truth created by humans. 

ROUGE includes several metrics, each serving a unique purpose in evaluating text. Two of the key metrics are ROUGE-N and ROUGE-L. ROUGE-N assesses the overlap of $n$-grams between the machine-generated text and the reference, where $n$ is 1 and 2 in our experiments. ROUGE-L focuses on the longest common subsequence (LCS) between the LLM-generated $n$ and the reference.

ROUGE-N is based on the $n$-gram overlap between the machine-generated text and the reference as follows:
\begin{equation}
    \text{ROUGE-N} = \frac{\sum_{s \in \{\text{Reference}\}} \sum_{n\text{-gram} \in s} \text{Count}_{\text{match}}(n\text{-gram})}{\sum_{s \in \{\text{Reference}\}} \sum_{n\text{-gram} \in s} \text{Count}(n\text{-gram})}
\end{equation}
where $\text{Count}_{\text{match}}(n\text{-gram})$ is the count of $n$-grams in the machine-generated text that matches the ground truth. $\text{Count}(n\text{-gram})$ is the count of $n$-grams in the ground truth. On the other hand, ROUGE-L evaluates the LCS between the machine-generated text and the reference as follows:
\begin{equation}
    \text{ROUGE-L} = \frac{\sum_{s \in \{\text{Reference}\}} \text{LCS}(s, \text{Machine})}{\sum_{s \in \{\text{Reference}\}} \text{Length}(s)}
\end{equation}
where $\text{LCS}(s, \text{Machine})$ refers to the length of the LCS between the system-generated text and the reference $s$. Finally, $\text{Length}(s)$ is the length of the reference text. For both BLEU and ROUGE, the higher the score, the better. We implemented both metrics using HuggingFace \texttt{evaluate} library \citep{huggingface2024bleu,huggingface2024rouge}. Note that we evaluate the text from LLM's answer that is generated in a downloadable file, not from the text-based responses.

\subsubsection{Considerations}
Achieving high BLEU and ROUGE scores requires a significant overlap between the LLM's output and our ground truth data, where `overlap' means identical wording. These metrics do not assess meaning but only textual similarity. For example, the sentences `He is 25' and `He was born in 2000' would yield low scores despite conveying the same information. Therefore, we must ensure the LLM returns data in a specific format, which we also use in our ground truth.
To achieve this, we designed tasks that are largely deterministic (solvable by traditional software) and provided examples within the prompt to guide the LLM. In the future, we plan to explore fine-tuning an LLM, which could enhance user experience. However, this study focuses on feasibility, and fine-tuning is beyond its scope.

\subsection{Common tasks for forensic timeline analysis}
\label{sec:task_definition}
Given the considerations and in order to quantitatively evaluate the capabilities of an LLM, we selected the following four tasks:
\begin{enumerate}
    \item Running \texttt{grep} for specific terms, i.e., assess how well the LLM handles a straightforward task such as running \texttt{grep}.
    
    \item Rule-based anomaly detection, i.e., looking for patterns that could also be identified using rules, such as multiple failed login attempts, could mean a brute-force attack. 
    
    \item Event summarization, i.e., combining several low-level events into a more meaningful event, such as if events A and B are found (low-level), this means a new user was created (meaningful event). 
    
    \item Exploratory data analysis.
\end{enumerate}

Tasks have been carefully chosen to be realistic but also allow for validation, e.g., for running grep we can develop our own grep expression. Note that only the first three tasks require a ground truth. 
With respect to the prompts, we follow the prompt style of \citet{scanlon2023digital} and the OpenAI prompt engineering guides \citep{openai2024prompt}. More details are provided in the subsequent sections.

\subsubsection{Prompts for running grep of specific terms}
This task simulates a simple \texttt{grep} command to ensure that it can handle basic tasks without making critical errors. The example prompts are shown below:

\begin{enumerate}
    \item ``I am a forensic investigator. I need to find these terms: \texttt{\textbackslash b[A-Za-z0-9\_\-\textbackslash\textbackslash :.]+\textbackslash .exe\textbackslash b} in the given CSV file to get all entries related to executable files (.exe). The CSV file is a forensic timeline generated from the log2timeline/Plaso tool.''
    \item ``For your references, the grep command is: grep -E ``\texttt{\textbackslash b[A-Za-z0-9\_\-\textbackslash\textbackslash :.]+\textbackslash .exe\textbackslash b}'' timeline.csv.''
    \item ``Do not include the first line of the file containing column names. Include all columns in the results, not only the message column. Export the results into plain text.''
\end{enumerate}

The prompt asks an LLM to replicate the functionality of a \texttt{grep} command, which is commonly used to search for patterns in the text. The goal is to search the CSV file for all entries that contain executable files with the \texttt{.exe} extension. In total, five terms need to be found, the system is expected to identify these entries and save the results in plain text format. In addition, we ask the system to exclude the header row and include all column values in the results. This task checks whether the LLM can effectively search through the forensic timeline using a regular expression to filter out relevant entries.

\subsubsection{Prompts for rule-based anomaly detection}

The goal of this task is to enable more natural queries against the timeline. This simulates providing a timeline and then asking about specific aspects, such as `Have there been failed login attempts?' or `Was registry.exe executed?'
Rather than posing these queries one by one, we opted to include multiple elements of interest in a file (keyword list), which the user uploads. This approach effectively cross-references a keyword list with the CSV-based timeline.

Specifically, we provide the following prompt: `I am a forensic investigator. Read this list of keywords to find suspicious events.' The user uploads a keyword list, allowing the system to focus on specific patterns or terms that may indicate abnormal or anomalous behavior within the timeline.

As we require the output in a specific format, the uploaded file is in reality a JSON file which includes elements of the prompt (\texttt{event}) as well as what to look for (\texttt{keyword}). This helps the LLM to detect suspicious events in a timeline CSV file. Note, the keyword is extracted from the message column from the timeline data, i.e., it exists in the timeline CSV. The event is our own creation. 

{\footnotesize
\begin{verbatim}
{
    "event": "Registry launch with prefetch file",
    "keyword": "Prefetch [REGEDIT.EXE] was executed"
}
\end{verbatim}
}

The LLM is expected to return a JSON-formatted response that includes the timestamp of the detected event (datetime), the name of the matched event (event), the keyword that triggered the match (keyword), and the full log message (message) from the timeline. This structured format facilitates automated comparison with ground truth data and supports downstream forensic analysis. This format also maintains consistency and interpretability to allow for an accurate evaluation using BLEU and ROUGE. An example output structure is shown below:

{\footnotesize
\begin{verbatim} 
[{
    "datetime": "2023-12-27T00:37:14.609465+00:00",
    "event": "Registry launch with prefetch file",
    "keyword": "Prefetch [REGEDIT.EXE] was executed",
    "message": "Prefetch [REGEDIT.EXE] was executed - 
    run count 3 path hints: \\WINDOWS\\REGEDIT.EXE 
    hash: 0x246AC210 volume: 1 [serial number: 
    0x5CE1DF5A  device path: \\VOLUME{01da182ce1985a
    64-5ce1df5a}]"
}]
\end{verbatim}
}

\subsubsection{Prompts for event summarization}

A user action (high-level event) of causes many entries in a timeline (low-level events). This tasks looks at the possibility to summarize low-level tasks to high-level tasks. To solve this task without fine-tuning, we provide a code (a python library) that can be used (executed) by the LLM.

The interaction between the user and ChatGPT is outlined in Fig.~\ref{fig:example-chat}. We provide a persona, such as stating a role (e.g., forensic investigator), including detailed information about the task (e.g., event type or data format), and offer additional tools to improve accuracy. These steps help the system to manage responses more accurately. The prompt uses a space delimiter to provide suitable spacing to separate key pieces of information.

In the third-to-last box (``Specify steps to run an event summarization''), the user outlines the exact procedure for summarizing events. This involves uploading the CSV file, selecting the type of event (such as ``last-shutdown''), and executing the summarization using the given libraries. The expected return value for this task is as follows:

\begin{figure*}[t]
	\centering
	\includegraphics[width=0.68\textwidth]{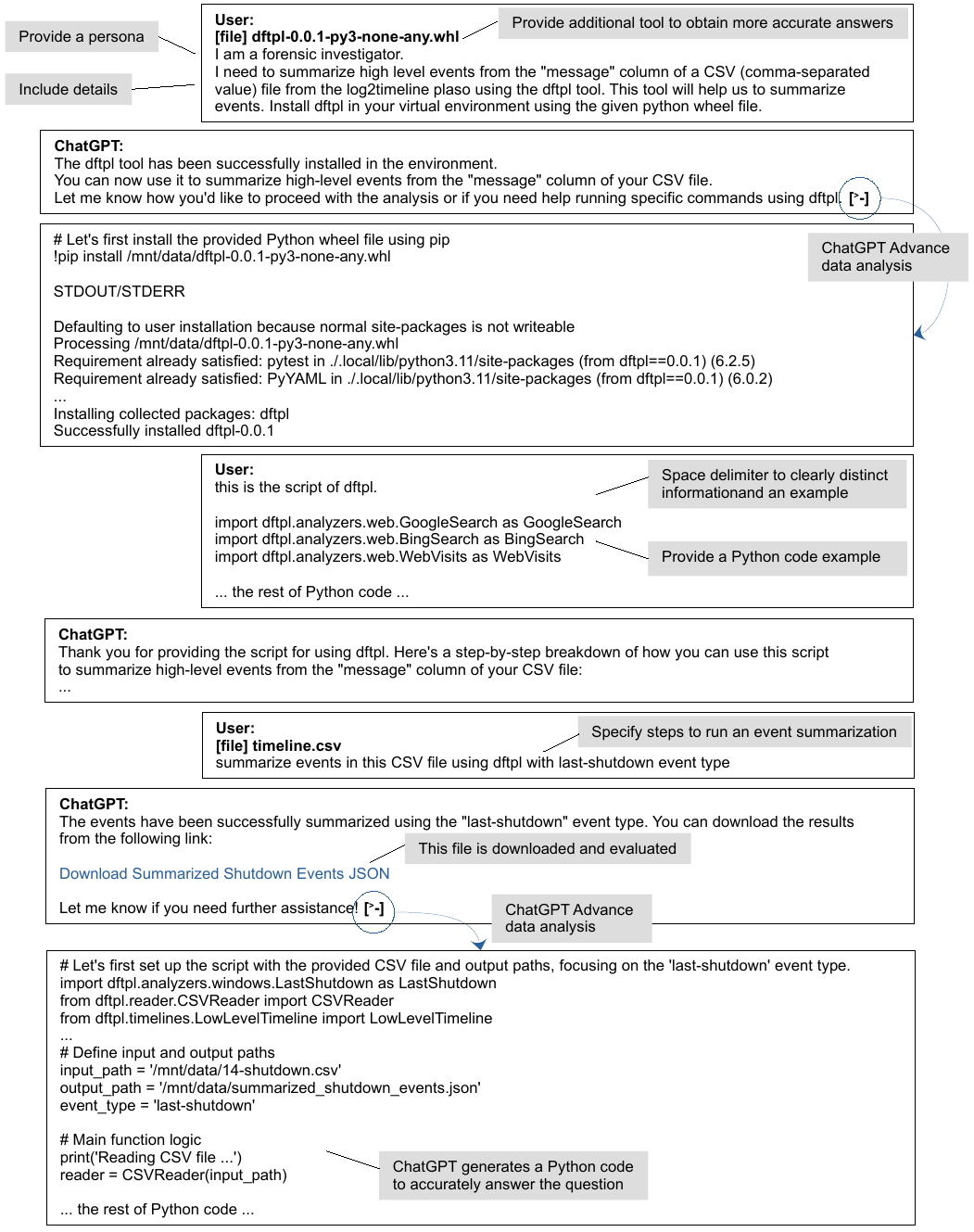}
	\caption{A conversation sample between an investigator and ChatGPT. Note that the prompt should be accurate following the OpenAI prompt engineering guides}
	\label{fig:example-chat}
\end{figure*}

{\footnotesize
\begin{verbatim}
{"0": {
 "id": 1002,
 "date_time_min": "2023-12-26 00:34:47.890403+00:00",
 "date_time_max": "2023-12-26 00:34:47.890403+00:00",
 "evidence_source": "[9707 / 0x25eb] Provider identifier: 
   {...} Source Name: Microsoft-Windows-Shell-Core Strings: 
   ['msedge.exe\" --no-startup-window --win-session-start'] 
   Computer Name: WinDev2311Eval Record Number: 2249 
   Event Level: 4",
 "type": "Process Creation",
 "description": "Process creation of 'msedge.exe'",
 "category": "Windows",
 "plugin": "EVT-WinEVTX-winevtx",
 "files": "NTFS:\\Windows\\System32\\winevt\\Logs\\
   Microsoft-Windows-Shell-Core%4Operational.evtx",
 "keys": {
   "Windows Event ID": "9707",
   "Windows Event ID (hex)": "0x25eb",
   "Executable name": "msedge.exe"
 },
 "supporting": { ... },
 "trigger": { ... }
} ... }
\end{verbatim}
}

\subsubsection{Prompts for exploratory data analysis}
Lastly, we explore the potential of LLMs for exploratory data analysis (EDA) which allows gaining valuable insights into the dataset as a whole. For instance, EDA may help investigators grasp the structure, distribution, and key features. It may also enable the identification of patterns and relationships between events, such as how user behaviors might be interconnected.
In addition, it facilitates the visualization of temporal data, which is an important aspect of timeline analysis. Using diagrams such as histograms and heatmaps, investigators can acquire a clearer understanding of trends and cycles in the data. These visualizations pinpoint periods of interest and aid in the identification of suspicious activities for further investigation. 

The example EDA prompt is: ``Explore patterns of event occurrences based on the datetime field per second (e.g., busiest times, significant gaps), use a bar chart. Write the hour:minute:second in the \textit{x} axis''. An LLM will generate a Python code to create the bar chart, and we can download the chart as a PNG file.

\subsection{Ground truth}
\label{sec:ground_truth}
To assess the quality of output (LLM response), we require a ground truth dataset, i.e., documentation of the underlying dataset \citep{gobel2023data,breitinger2023sharing}. A peculiarity in our scenario is that we need the ground truth in a specific format so that it is comparable with the output of an LLM (automated). Specifically, there is no easy way to compare a disk image or its corresponding timeline against the LLM output. Consequently, the underlying dataset must be converted into a text-based format (ground truth), allowing and automated comparison with the LLM output. 

To accomplish this, we first must create a dataset (Sec.~\ref{sec:building_a_dataset}) where the creation process is documented or recorded. Next, we generate a timeline of the disk image (Sec.~\ref{sec:timeline_generation}) which serves as an input for the LLM. Lastly, using the documentation and timeline, we manually create the \emph{expected outcome} which represents our ground truth (Sec.~\ref{subsubsec:event-reconstruction} to \ref{sec:ground_truth_grep}).

\subsubsection{Scenario and dataset generation}\label{sec:building_a_dataset}
\begin{figure}[th]
	\centering
	\includegraphics[width=0.4\textwidth]{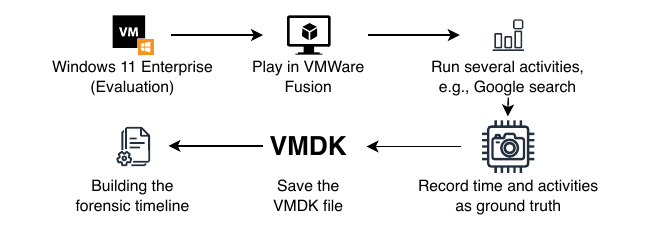}
	\caption{Building ground truth for LLM evaluation}
	\label{fig:building-ground-truth}
\end{figure}

The first step was to create a dataset as no appropriate dataset was available. The procedure is illustrated in Fig.~\ref{fig:building-ground-truth} and the dataset is shared through Zenodo. 
Our test bed was a Windows 11 machine within a virtual environment simulating regular computer usage. All activities were recorded using screen capture (video) and therefore are documented (written notes). 

The scenario follows a sequence of opening applications, downloading software, and accessing websites. The user begins by opening the Edge browser and then navigates to Bing. They perform a search query for ``Mozilla Firefox download'' on Bing and visit Mozilla’s official website to download the Firefox browser. After that, the user opens the File Explorer to navigate the downloaded installer. The user runs the Firefox installer and opens the newly installed Firefox browser. Afterward, they navigate to Google, perform a search related to SQL injection, and open a tutorial on the W3Schools website. The session ends with a system shutdown, indicating that the user has completed all activities.

\subsubsection{Timeline generation}\label{sec:timeline_generation}
To generate the timeline, we ran log2timeline/Plaso \citep{metz2024log2timeline} on the \texttt{vmdk} file. 
The tool (Plaso) analyzes all known artifacts\footnote{Plaso consists of various parsers for different artifacts. Artifacts unknown to Plaso are ignored.} and compiles them into a single unified timeline. 
By default, the tool processes all partitions from a \texttt{vmdk} file and generates a Plaso storage file (\texttt{*.plaso}, a database file) containing the forensic timeline.
To convert \texttt{plaso} file to a CSV timeline file, we ran \texttt{psort}.

\subsubsection{Ground truth for Task 1: Running \texttt{grep} for specific terms}
\label{sec:ground_truth_grep}
Building the ground truth is straightforward as we only have to manually run \texttt{grep} on the dataset and take note of the output. This was done for the following five keywords:

\begin{enumerate}
    \item \texttt{RegisteredApplications}: obtaining events related registered applications in Windows registry.
    \item \texttt{(OneDrive|OneDrive\textbackslash.exe)}: finding events related to Microsoft OneDrive application.
    \item \texttt{\textbackslash b[A-Za-z0-9\_\-\textbackslash\textbackslash :.]+\textbackslash .exe\textbackslash b}: looking for all entries related to executable files (.exe).
    \item \texttt{4616 /}: finding  Windows event ID 4616 which related to system time change without regex.
    \item \texttt{\textbackslash[4616 / 0x1208\textbackslash].*Microsoft-Windows-Security}
    \texttt{-Auditing.*svchost.exe}: finding Windows event ID 4616 with regex. 
\end{enumerate}

The command to generate this ground truth is \texttt{grep -E keyword timeline.csv}, where \texttt{-E} indicates that extended regular expressions are being used with the \texttt{grep} command.

\subsubsection{Ground truth for Task 2: Rule-based anomaly detection}
\label{subsubsec:keyword-analysis}
The second ground truth requires matching keywords (or phrases) with events. 
We create the keywords as a rule-based approach by first checking the date and time of the event we performed earlier in the Windows test-bed. Next, we manually look for related entries in the timeline CSV file. Once we find the relevant entry, such as registry launch, we extract the keywords linked to the event. Finally, we format these keywords into a JSON format as shown below:
{\small
\begin{verbatim}
{
    "event": "Registry launch with prefetch file",
    "keyword": "Prefetch [REGEDIT.EXE] was executed"
}
\end{verbatim}
}
In the evaluation, we can ask questions in natural language because the event and the keyword the LLM searches for are already defined. Unlike an event summarization task, no script or library is provided, and the LLM handles the matching on its own.
These keywords collected are a useful technique to identify suspicious events in the forensic timeline. There are seven keywords in total and the full list of keywords in JSON format is available on Zenodo.

\subsubsection{Ground truth for Task 3: Event summarization}
\label{subsubsec:event-reconstruction}
Event summarization aims at combining low-level events to obtain high-level events based as proposed by \citet{Hargreaves2012}. Forming the ground truth was accomplished by implementing the \texttt{dftpl} tool\footnote{\url{https://github.com/studiawan/dftpl}} as described by the authors.
Given a CSV timeline, our prototype can extract certain high-level events and return a JSON file. 
There are eight predefined events, grouped into three categories:
\begin{enumerate}
    \item Web: Google search, Bing search, and web visit
    \item Windows: last shutdown, process creation, and program opened
    \item User activity: file download, and recent file access
\end{enumerate}

We chose JSON because it is human-readable, making it easier for investigators to interpret and manually validate results. JSON also facilitates straightforward comparison with evaluation metrics due to its structured nature for efficient parsing. Moreover, its compatibility with various programming languages and tools further supports automation and quantitative evaluation in forensic analysis workflows.

To create the high-level events, we ran the \texttt{dftpl} command as follows: \texttt{dftpl -i timeline-input.csv -o summarization-output.json -t last-shutdown}, where 
\texttt{-i} is a Plaso CSV file,
\texttt{-o} is the output (in JSON), and
\texttt{-t} specifies the event of interest. The \texttt{-t} option can be omitted to summarize multiple events. 
The list of high-level events was then manually validated that it was correct.

\begin{figure*}[!t]
	\centering
	\includegraphics[width=0.7\textwidth]{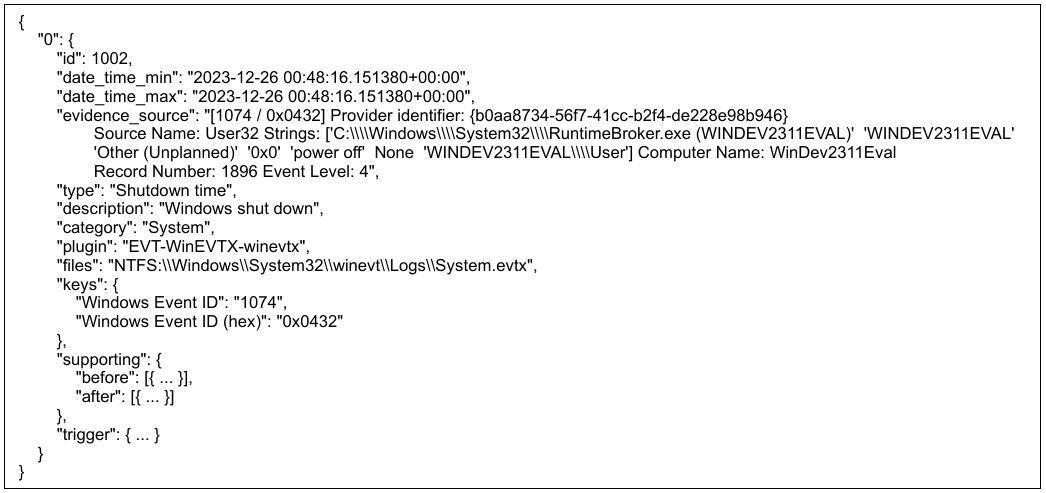}
	\caption{An example of a ground truth for the event summarization task in JSON format}
	\label{fig:ground-truth-sample}
\end{figure*}

A sample output is provided in Fig.~\ref{fig:ground-truth-sample} and includes the following high-level activities:
\begin{enumerate}
    \item \textit{id}: A unique identifier for the event, which is a number that differentiates this event from others.
    \item \textit{date\_time\_min}: The earliest possible timestamp for when the event could have occurred.
    \item \textit{date\_time\_max}: The latest possible timestamp for when the event could have occurred.
    \item \textit{evidence\_source}: Refers to the Plaso message that provides information about the event.
    \item \textit{type}: The nature of the event, such as Google Search, File Download, or any other high-level event type.
    \item \textit{description}: A human-readable explanation or summary of the event.
    \item \textit{category}: A higher-level classification or tag for filtering or organizing events.
    \item \textit{plugin}: Identifies the Plaso plugin used to parse the source file from which the event was extracted.
    \item \textit{files}: Refers to the file(s) related to the event, such as the log file, binary file, or any other data source.
    \item \textit{keys}: Stores additional key-value pairs related to the event, such as specific attributes or metadata. 
    \item \textit{supporting}: Stores a list of five low-level events before and after the main event for context.
    \item \textit{trigger}: Refers to the reasoning artifact or piece of evidence that caused the event to be recognized. 
\end{enumerate}

\section{Experimental results and analyses}
\label{sec:experiments}

This section details the experimental settings, along with the analysis, results, and discussion of our case study.

\subsection{Experimental settings}

We used the version of log2timeline/Plaso which was the Docker image version 20230717. The target operating system was Microsoft Windows 11 Enterprise. The OS was sourced from the Microsoft Developer Network, specifically the evaluation virtual machine (VM) version 2311 \citep{microsoft2024get}. For virtualization, we opted for VMWare Fusion 13.5.0.
For the LLM, we selected ChatGPT-4o, one of the most advanced models available at the time of writing this paper. To facilitate containerized environments, we use Docker Desktop version 4.22.1 (118664). 

The extracted full timeline is too large to be handled by ChatGPT due to token limitations. Consequently, we only provided ChatGPT with about 2000 lines of Plaso entries as a timeline of interest. We have experimented with different sizes (e.g., 1000, 2000, 3000 lines) and found 2000 lines to be a manageable amount that balances input size and processing efficiency.

\subsection{Timeline analysis with ChatGPT}
\label{subsec:timeline-analysis}
The Advanced Data Analysis feature of ChatGPT, previously called Code Interpreter, is now integrated into ChatGPT versions 4 and 4o \citep{openai2024data}. This feature allows users to analyze data and interpret code directly within the platform. This enhances the user experience by supporting data uploads, where users can write, test, and execute code seamlessly.
The supported file formats include text, image files, PDFs and Word documents, code or other data files, as well as audio and video. In this study, we used the CSV file generated by Plaso. Once the data is uploaded, we can use the prompts to instruct ChatGPT to read or analyze the timeline.

We employ ChatGPT in two scenarios: with and without additional knowledge. In the first scenario, we provided ChatGPT with specific information related to the task, such as a library for event summarization (Sec.~\ref{subsubsec:event-reconstruction}) or a list of keywords to detect suspicious activities (Sec.~\ref{subsubsec:keyword-analysis}). In the latter scenario, we did not provide any additional information and relied solely on ChatGPT's existing language model to analyze the timeline.

\subsection{Results and analysis}

To quantitatively evaluate ChatGPT for forensic timeline analysis, we developed four tasks, including ground-truth data. For example, the event summarization task has 14 event types, the rule-based anomaly detection task has seven rules, and the search task for specific terms has five keywords. 
Note that the exploratory data analysis task does not have evaluation metrics because there is no ground truth data for this task.

A sample result of the given prompts and the ChatGPT answers is depicted in Fig.~\ref{fig:example-chat}. 
The evaluation results for the used datasets are shown in Table~\ref{tab:experimental-results} where the metric values represent the mean values for each task.

\begin{table*}[t!]
\centering
\footnotesize
\caption{Evaluation results of various tasks given to ChatGPT for forensic timeline analysis}
\label{tab:experimental-results}
\begin{tabular}{lrrrrr}
\toprule
Task                                   & BLEU  & ROUGE-1 & ROUGE-2 & ROUGE-L & Mean score\\ 
\midrule
\textit{Without additional knowledge} & & & & & \\
Event summarization (single)          & 0.077 & 0.192   & 0.129   & 0.136   & 0.134 \\
Event summarization (multiple)        & 0.001 & 0.171   & 0.120   & 0.132   & 0.106 \\
Rule-based anomaly detection          & 0.147 & 0.144	& 0.075   & 0.141   & 0.127 \\
Run grep for specific terms           & 0.847 & 1.000   & 1.000   & 1.000   & 0.962 \\ 
\midrule
\textit{With additional knowledge} & & & & & \\
Event summarization (single)          & 0.999 & 1.000   & 1.000   & 1.000   & 1.000 \\
Event summarization (multiple)        & 0.743 & 0.786   & 0.786   & 0.786   & 0.775 \\
Rule-based and anomaly detection      & 0.945 & 0.997   & 0.996   & 0.997   & 0.984 \\
Run grep for specific terms           & 0.847 & 1.000   & 1.000   & 1.000   & 0.962 \\ 
\bottomrule

\end{tabular}
\end{table*}

\subsubsection{Results of running grep for specific terms}

It is important to note that when asked to search for specific terms, ChatGPT does not run the \texttt{grep} command. Instead, it generates Python code to perform the search. The results of this task are shown in Table~\ref{tab:experimental-results}. 
The results indicate that ChatGPT performs this task effectively, especially when provided with additional knowledge, i.e., the corresponding \texttt{grep} command. 
Without additional knowledge, the BLEU score is 0.847, and both ROUGE-1 and ROUGE-L are 1.000. The results suggest that the system accurately identifies specific terms most of the time, but with minor variations that affect the BLEU score. With additional knowledge, the BLEU, ROUGE-1, and ROUGE-L scores all reach 1.000 and they demonstrate that the model can perfectly match the specific terms when it has more context or knowledge about the data. These findings imply that the performance of ChatGPT in conducting targeted searches is enhanced when it is given relevant prior information. Therefore, it produces consistent and fully accurate results. 

ChatGPT can detect all entries correctly when provided with additional knowledge or information. However, the \texttt{grep} output from ChatGPT does not contain commas, whereas the ground truth does, as the timeline is a comma-separated file. Additionally, the model's output has extra spaces that are not present in the original data.

Furthermore, it gives inconsistent output when no additional knowledge is provided. In several cases, it only produces incomplete results, displaying only the ``message'' column without including all other columns. In other instances, it provides the correct values for all columns of the CSV file. When we obtained inconsistent responses, we clicked the ``Refresh'' button and it would generate the correct ones.

\begin{figure}[t!]
	\centering
	\includegraphics[width=0.5\textwidth]{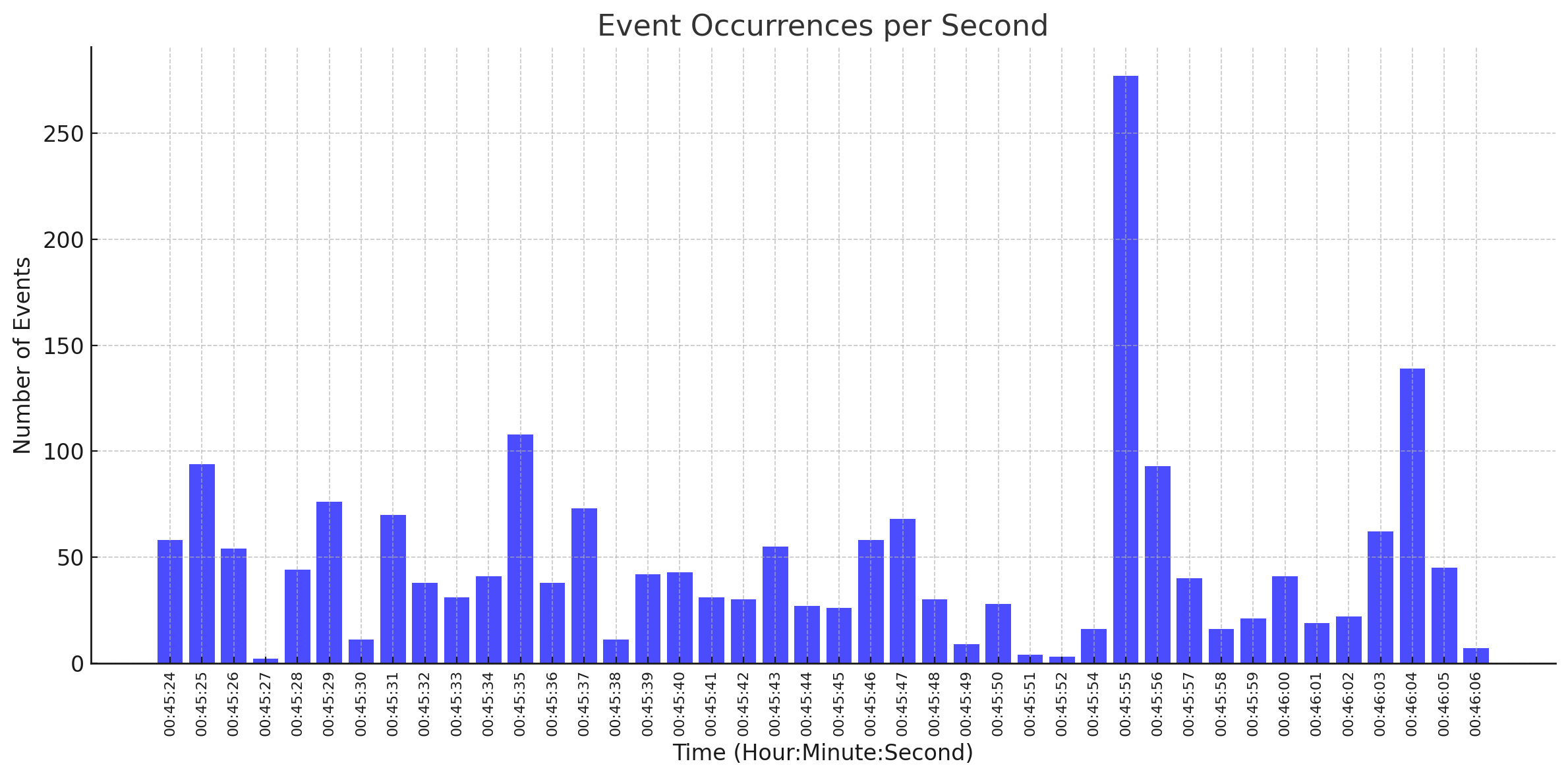}
	\caption{A bar chart generated by ChatGPT in exploratory data analysis task.}
	\label{fig:bar-chart}
\end{figure}

\subsubsection{Results of rule-based anomaly detection}

As mentioned in Sec.~\ref{subsec:timeline-analysis}, there are two scenarios: one with additional knowledge and one without. In the case without additional knowledge, the prompt is slightly different because it does not include instructions to read the uploaded keywords file. In this task, we can instruct ChatGPT to format the answers in a specific format, such as JSON. The prompt would be ``Format your answer using this JSON format:'' and we can give an example format as follows:

{\footnotesize
\begin{verbatim}
{
  "datetime": "datetime_here",
  "event": "event_name_here",
  "keyword": "keyword_here",
  "message": "message_from_logs_here"
}
\end{verbatim}
}
Moreover, we instruct the system to export all results to a downloadable file, with ``I need all entries of suspicious entries. Export to a JSON file for all of the results''. 

In the task of rule-based anomaly detection without additional knowledge, the performance was poor: 
The BLEU score is 0.147, and the ROUGE scores range from 0.141 to 0.192, indicating that the model's output is different significantly from the expected output. 
The keywords generated by ChatGPT are as follows: `delete', `clear', `wipe', `remove', `malware', and `unauthorized'. These low scores reflect minimal overlap between the system's output and the expected results, both in terms of individual words and word sequences. However, it is important to note that these evaluation metrics are based on word matching and do not account for semantic similarity. Although the wording used by ChatGPT may differ from the predefined ground truth, the underlying interpretation or intent of the result may still be forensically relevant or correct.

In contrast, the results improve when additional knowledge is provided. Specifically, the BLEU score rises to 0.945 and the ROUGE scores increase to nearly perfect values (ranging from 0.996 to 0.997). 
This means that generated outputs closely match the expected results. This highlights the importance of providing context or specialized knowledge to improve performance in more complex forensic analysis tasks.

Even with additional information or knowledge, ChatGPT can still make mistakes. The errors are mainly due to differences in how characters are escaped. For example, the ground truth uses two backslashes to escape regular expressions (regex), while ChatGPT's output uses four backslashes to escape the ``\textbackslash'' character.

\subsubsection{Results of event summarization}
\label{subsubsec:event-reconstruction-results}

Event summarization comprises two scenarios: summarizing a single event or multiple events. 
Summarizing a single event means the method extracts one specific event from the provided timeline, such as a Google search (full list see Sec.~\ref{subsubsec:event-reconstruction}). Consequently, multiple events mean the LLM is tasked with summarizing all defined events. 

Our research indicates that ChatGPT uses a virtual environment to run Python code when responding to user prompts. This means we can install the \texttt{dftpl} Python wheel installer within that virtual environment. 
To respond to the user prompts, ChatGPT generates Python code as shown in Fig.~\ref{fig:example-chat}. For example, if the parser example is designed to work for all supported events, ChatGPT can summarize a specific event, such as the last shutdown event on Windows. One can click the `[$>$\_]'-button to view the generated Python source code. Thus, experienced investigators may validate the code and with it the answer. Finally, the results can be downloaded in a JSON format and this file will be quantitatively evaluated based on the ground truth from Sec.~\ref{subsubsec:event-reconstruction}.

The result of event summarization on single and multiple events without additional knowledge shows a low performance, with a BLEU score of 0.077 indicating limited precision in generating a summarization that closely matches the expected events. The ROUGE-1 score of 0.192 suggests that around 19.2\% of single words in the generated output matched the reference, while the ROUGE-2 score of 0.129 shows even lower overlap in bigrams (two-word sequences). The ROUGE-L score of 0.136 reflects a moderate match in terms of the longest sequence of matching words. However, we conclude that without additional knowledge, the system cannot accurately summarize events.

In contrast, the result for a single event with additional knowledge, i.e., using the \texttt{dftpl} library, shows near-perfect performance, with a BLEU score of 0.999 and ROUGE-1, ROUGE-2, and ROUGE-L scores, all at 1.000. This indicates that the ChatGPT output almost exactly matched the reference in terms of precision, word overlap, and sequence structure. The high scores suggest that, with additional knowledge, the system was able to mimic the expected results. The reason is that we gave a Python library that can summarize events based on the method described in \citet{Hargreaves2012} to ChatGPT (Fig.~\ref{fig:example-chat}). Although we did not explicitly instruct ChatGPT to follow a particular order, the ground truth output produced by the \texttt{dftpl} library is chronologically ordered by timestamp.
For the multiple event summarization task, the evaluation scores were lower because ChatGPT generated the correct events but in a different order than the ground truth. 
The beginning of the file displays timestamps that increase or remain the same, indicating a mostly sorted order. Similarly, the end of the file follows a chronological pattern. However, the middle sections break this order, with some events appearing earlier than preceding ones.
This discrepancy in ordering affected the BLEU and ROUGE scores, which are sensitive to the sequence of words or structures. Importantly, while the order differed, the extracted content was sometimes semantically correct and forensically valid. Future work may include implementing order-invariant evaluation metrics or normalizing the output order before comparison to address this issue.

\subsubsection{Results of exploratory data analysis}

This section aims to explore how ChatGPT can assist forensic investigators in identifying patterns or anomalies within large timelines through exploratory data analysis (EDA). Specifically, we evaluate the model's ability to generate useful visualizations that support investigative tasks.
The example of a generated bar chart is shown in Fig.~\ref{fig:bar-chart}. The chart shows the number of events occurrences per second within a specific time range, where each bar corresponds to a second in the format: \textit{hour:minute:second}. The data reveal variability in event activity, with most seconds seeing between 50 and 150 events. However, there is a noticeable spike at 00:45:55, where the event count exceeds 250 which indicates a sudden surge in activity during that particular second. The concentration of events at specific seconds may point to important actions or incidents that require further investigation, especially during periods of relatively low activity that are punctuated by intense bursts \citep{Studiawan2021}. 

\begin{figure}[t!]
	\centering
	\includegraphics[width=0.5\textwidth]{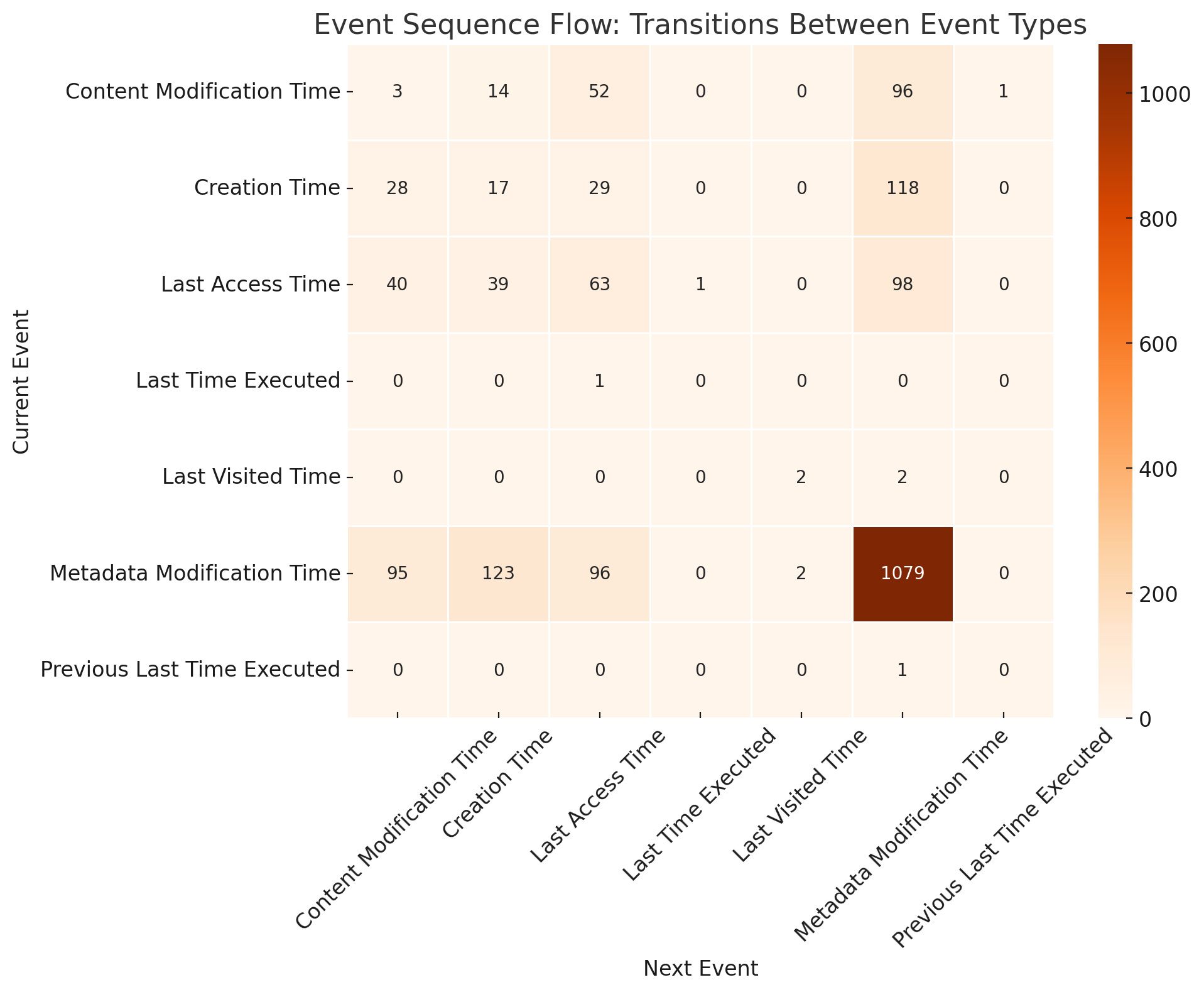}
	\caption{A heatmap generated by ChatGPT for event sequence flow}
	\label{fig:heatmap-sequence-flow}
\end{figure}

Another chart generated by ChatGPT using Python is a heatmap shown in Fig.~\ref{fig:heatmap-sequence-flow}. The heatmap illustrates the flow of the event sequence, showing the transitions between various types of events based on their timestamps. The rows represent the current event types, while the columns represent the next event types, with each cell indicating how often a specific event type is followed by another. The color intensity, as shown by the legend, reflects the frequency of these transitions, with darker shades showing more frequent sequences. 
The heatmap highlights common flows in the event timeline and provides valuable insight into which events tend to trigger others. Therefore, it can help to understand the sequences of events within the forensic timeline analysis. 

Key patterns can be observed in this visualization. For example, `Metadata Modification Time' transitions into itself 1079 times, suggesting that it frequently repeats or is followed by itself in the sequence of events. There are also noticeable transitions from `Creation Time' to `Metadata Modification Time' (118 times) and from `Last Access Time' to `Metadata Modification Time' (98 times). 

The heatmap reveals typical patterns in the event timeline by showing how certain events frequently follow others. This visualization helps investigators better understand the sequence and relationship between events during forensic timeline analysis.
In short, EDA can be done by a human investigator, but using ChatGPT can help speed up this manual work.

\subsection{Discussion}

\paragraph{Overall quantitative evaluation}

Without additional knowledge, tasks such as `Event summarization (single)' and `rule-based anomaly detection' have mean scores of 0.134 and 0.127, respectively, indicating limited accuracy. However, `run grep for specific terms' achieves a much higher mean score of 0.962, suggesting that ChatGPT can handle search for specific terms relatively well even without prior information. With additional knowledge, mean scores improve across all tasks. Single event summarization tasks achieved a perfect mean score of 1.000, while the multiple one obtained 0.775. 
The results demonstrate inconsistent accuracy scores, even when provided with relevant context. The mean score for the rule-based anomaly detection task also increases to 0.984. The consistent mean score of 0.962 for ``run grep for specific terms'' shows that the task is already handled effectively regardless of additional knowledge. In the \texttt{grep} task, providing prior information does not lead to further improvement.

\paragraph{CSV file size of a forensic timeline}
While ChatGPT is advertised as being capable of handling CSV files up to 50MB in size\footnote{\url{https://help.openai.com/en/articles/8983719-what-are-the-file-upload-size-restrictions}}, we found that in practice, it struggles to process files of that size. Throughout our experiments, we observed that ChatGPT could successfully analyze smaller CSV files, but when attempting to work with larger files (10MB or more), the model often encountered errors or failed to provide results. This discrepancy suggests that, despite the claims in the documentation, there are practical limitations when analyzing larger datasets, likely due to resource constraints or the tokens complexity involved in processing such large volumes of data.

\section{Conclusion and future work}
\label{sec:conclusion}

The proposed methodology and dataset have demonstrated its potential for quantitative evaluation of timeline analysis using LLMs. Using the proposed standardized methodology and dataset,  researchers can apply and expand the test and evaluation of LLM-based forensic timeline analysis. By employing the advantages of natural language processing on LLMs, e.g., ChatGPT, the presented case studies show that it can assist in analyzing events and temporal information from a forensic timeline. It also provides valuable information for forensic investigators, particularly in the task of exploratory data analysis. However, based on the quantitative evaluation, ChatGPT performs worse than a rule-based approach or a regular expression-based approach accompanied by a human investigator.

For future work, we plan to undertake the task of developing custom LLMs specifically trained on digital forensic data and associated tasks. In addition, we will explore the use of other commercial LLM services such as Google Gemini and Claude to evaluate the robustness of our approach. In addition, to address concerns about the confidentiality of digital evidence, we plan to deploy open-source LLMs, such as LLaMA \citep{touvron2023llama} and Mixtral \citep{jiang2024mixtral} on a local device. By keeping the forensic timeline on the local computer, we aim to avoid the need to upload sensitive data to cloud-based LLM services, thus ensuring the privacy of the investigation.

\section*{Acknowledgments}
We would like to thank Christopher Hargreaves for his valuable comments and feedback.

\balance

\bibliographystyle{elsarticle-harv} 
\bibliography{mybibfile}

\end{document}